\documentstyle[preprint,epsf,aps]{revtex}
\begin{document}
\title{Producing the event-ready  two photon polarization entangled  state
 with normal photon detectors}
\author{ Wang Xiang-Bin\thanks{email: wang$@$qci.jst.go.jp}\thanks{The initials of the author's name are X.B.} 
\\
        Imai Quantum Computation and Information project, ERATO, Japan Sci. and Tech. Corp.\\
Daini Hongo White Bldg. 201, 5-28-3, Hongo, Bunkyo, Tokyo 113-0033, Japan}

\maketitle 
\begin{abstract}
We propose a scheme to produce the maximally two photon polarization
  entangled state (EPR state) with  single
photon sources and the passive linear optics devices. In particular, our scheme
only requires the normal photon detectors which distinguish the vacuum and non-vacuum 
Fock number states. A sophisticated photon detector  distinguishing 
one-photon state and two-photon state is unnecessary in the scheme.
\end{abstract}
The resource of maximally  entangled state (EPR state) plays a fundamentally 
important role in testing quantum laws
related to the non-locality\cite{ein} and in many tasks of quantum information 
processing\cite{qip,chuang} such as the quantum teleportation\cite{bennett,bou}, quantum dense
coding\cite{bennett}
the entanglement based quantum key distribution\cite{ekert} and quantum computation\cite{chuang,gc}.
So far, it is generally believed that the  two photon polarization EPR state is
particularly useful in quantum information processing.

In quantum teleportation, initially two remotely separated
parties Alice and Bob share an entangled pair of
particle 2 and 3. Alice is offered with particle 1 which is in  an unknown state
$|u\rangle$. Alice's task here is to produce the unknown state $|u\rangle$ in Bob's side
without sending particle 1 to Bob or taking any action to observe
the state information of particle 1. To do so, she takes a joint measurement on particle 1 and 2
in the Bell-state basis. By observing the measurement result she does not know any information of the original 
state of particle 1, however, she
 knows that by 
what unitary transformation the state
of particle 3 at Bob's side can be  transformed to the unknown state $|u\rangle$.
In such a task, the resource of pre-shared (event-ready) entanglement between Alice and Bob is
crucial for a deterministic quantum teleportation\cite{remark1} . In the quantum dense coding, with the help of pre-shared
event-ready entanglement, Alice may send 2 bits information to Bob by only sending
him one quantum bit (a two-level-state particle)\cite{bennett}.    

The observation of EPR pairs has been carried out by many experiments 
(for example, ref\cite{para}). However, those entangled polarized photon pairs
were only produced randomly among vacuum states since there is no way to know whether a
polarization EPR pair is generated without destroying the state itself. To make sure
which states are indeed  EPR states normally we have to observe them and destroy 
them. That is to say, we are sure which ones had been in  EPR states only after we have destroyed the states. 
We call such a case as post-selection entanglement.
The post-selection property in the EPR state generation is not a serious drawback
in some quantum tasks such as the testing of the violation of 
Bell inequality. However, in many other tasks, such as the
deterministic quantum teleportation and
 quantum dense coding\cite{bennett}, the resource of event-ready
entanglement is a must. (Event-ready EPR means that we are sure certain pair is in EPR state and
the state is not destroyed.)  

The study of event-ready entanglement can be dated back to 1993\cite{zuo}.
Recently, some proposals are raised to make the $event-ready$ two photon
polarization
EPR state or the photon number entangled state. 
(One may refer to Ref\cite{kokt} for a complete description on
the theoretical condition to produce event-ready EPR pair through the parametric down conversion.) 
Among all the proposals(see, e.g., 
Ref.\cite{kokt,hof,zou,nent}) to produce the
entangled states in either polarization space or the photon number space, 
most of them
demand both single photon sources and sophisticated photon detectors which
can distinguish one-photon Fock state and two-photon Fock state. Both the 
single photon sources and the the sophisticated photon detectors are difficult
techniques and they are thought to be the main barriers to produce the
event-ready polarization EPR pairs with currently existing technology in
linear optics\cite{zou}. 
Although both the single photon source\cite{single1,single2,single3,single4,single5} and the imperfect 
sophisticated photon detector\cite{det} have been demonstrated already however, it is generally believed
that both of them are rather difficult technilogies. On the other hand, so far
 a successful $combination$ of these two techniques in one experiment
 has never been reported. Therefore it should be interesting to seek new schemes
which do not depend on either of these two sophisticated techniques. 
Very recently, Sliwa and Banaszek\cite{bana} proposed a scheme not demanding the single photon source
but still demanding a sophisticated photon detector. So far, all 
proposals for event-ready polarizatin EPR pairs with passive linear optics devices depend on the sophisticated photons
detectors to distinguish one-photon state and two-photon state, although normal photon detectors
with very good detection efficiency are in principle enough for producing the event-ready entanglement
in photon number space\cite{eisert0,eisert}. Normally,
a sophisticated detector cannot be replaced by a cascaded system of
many normal photon detectors unless the efficiency of the normal photon detectors are impractically high\cite{kb}. 
 In this paper, we propose a totally new scheme. 
Our new scheme requires the single photon sources but 
only uses normal photon detectors which only distinguish the vacuum and 
non-vacuum Fock number states. Moreover, our result is insensitive to detection efficiency of those photon detectors.

Our scheme is schematically shown in figure 1. 
In this scheme, our task is to observe the  following coincident event:
\\{\bf Coincidence}: {\it Both  detectors} D$_3$ $and$ D$_2$ {\it are clicked;} or
{\it both} D$_1$ {\it and} D$_4$ {\it are clicked.}\\ 
 If the above coincident event is observed, then we believe
that beam 2' and 4' are in the singlet EPR
state: 
\begin{eqnarray}|\Psi^-\rangle_{2'4'}
=\frac{1}{\sqrt 2}(|H\rangle_{2'}|V\rangle_{4'}-|V\rangle_{2'}|H\rangle_{4'})
\end{eqnarray}
The 4 input beams, beam 1,2,3 and 4 are from single photon sources. 
The polarization of both beam 1 and beam 3 deviate a little
bit from the vertical polarization while polarization of both
  beam 2 and beam 4 deviate a little
bit from the horizontal one. 
We assume that
we have tuned the optical paths very carefully so that to each beam splitter (BS) or polarizing beam splitter (PBS), 
the two input beams reach it simultaneously. A detailed study on the time window related to the quantum coherence had been  shown in the seminal work by Zukowski, 
Zeilinger, Horne and Ekert\cite{zuo}. 
A further study can be seen e.g., in \cite{zuo1}.  

Mathematically, the total input state is
\begin{eqnarray}
|V'\rangle_1|H'\rangle_2|V'\rangle_3|H'\rangle_4,
\end{eqnarray}
where the subscripts indicate the different beams,
$|H'\rangle=\frac{1}{\sqrt{1+|\epsilon|^2}}(|H\rangle+\epsilon |V\rangle)$
and $|V'\rangle=\frac{1}{\sqrt{1+|\epsilon|^2}}(\epsilon |H\rangle- |V\rangle)$
, $|\epsilon|<< 1$ and $|H\rangle$ and $|V\rangle$ are horizontally and 
vertically polarized states respectively.
A PBS has the property to transmit the horizontal polarization and reflect the vertical polarization as shown in figure 2.
After the four input beams reached the two PBS (PBS1 and PBS2),  beams 1' and
2' are in the following state:
\begin{eqnarray}|\chi\rangle_{1'2'}=\frac{1}{1+|\epsilon|^2}
(\epsilon|H\rangle_{1'}-|V\rangle_{2'})(\epsilon |V\rangle_{1'}
+|H\rangle_{2'})
\end{eqnarray}  
Similarly, beam 3' and 4' are in the following state 
\begin{eqnarray}|\chi\rangle_{3'4'}=\frac{1}{1+|\epsilon|^2}
(\epsilon|H\rangle_{3'}-|V\rangle_{4'})(\epsilon |V\rangle_{3'}
+|H\rangle_{4'}).
\end{eqnarray}
Beam 2' and 4' are now the outcome beams and they are a good entangled pair
if  1' and 3' are collapsed to the singlet state $|\Psi^-\rangle_{1'3'}$. This can be shown by mathematically 
recasting the  product state $|\chi\rangle_{1'2'}\otimes |\chi\rangle_{3'4'}$ . This product state is
\begin{eqnarray}
\frac{1}{(1+|\epsilon|^2)^2}\left[X_0+\epsilon X_1 + \epsilon^2(A+B+C)+O(\epsilon^3)\right]
\end{eqnarray}
where $X_0=|V\rangle_{2'}|H\rangle_{2'}|V\rangle_{4'}|H\rangle_{4'}$; \\
$X_1=(|V\rangle_{1'}|V\rangle_{2'}-|H\rangle_{1'}|H\rangle_{2'})
|V\rangle_{4'}|H\rangle_{4'}+(|V\rangle_{3'}|V\rangle_{4'}-|H\rangle_{3'}|H\rangle_{4'})|V\rangle_{2'}|H\rangle_{2'}$;\\
$A=|H\rangle_{1'}|V\rangle_{1'}|V\rangle_{4'}|H\rangle_{4'}$, $B=|V\rangle_{2'}|H\rangle_{2'}|H\rangle_{3'}|V\rangle_{3'}$ and
\begin{eqnarray}
C=(|H\rangle_{1'}|H\rangle_{2'}-|V\rangle_{1'}|V\rangle_{2'})
(|H\rangle_{3'}|H\rangle_{4'}-|V\rangle_{3'}|V\rangle_{4'}).
\end{eqnarray}
Moreover, $C$ is equivalent to
\begin{eqnarray}
C=|\Phi^+\rangle_{1'3'}|\Phi^+\rangle_{2'4'}+|\Phi^-\rangle_{1'3'}|\Phi^-\rangle_{2'4'}
-|\Psi^+\rangle_{1'3'}|\Psi^+\rangle_{2'4'}-|\Psi^-\rangle_{1'3'}|\Psi^-\rangle_{2'4'}
\end{eqnarray}
and $|\Phi^\pm\rangle_{ij}=\frac{1}{\sqrt 2}(|H\rangle_i|H\rangle_j\pm |V\rangle_i|V\rangle_j)$; 
$|\Psi^\pm\rangle_{ij}=\frac{1}{\sqrt 2}(|H\rangle_i|V\rangle_j\pm |V\rangle_i|H\rangle_j)$.
Therefore the rest of our job is to distinguish $|\Psi^-\rangle_{1'3'}$ from all other possible
states of of beam 1',3'. Note that in our case, all other possible states of beam 1', 3' are orthogonal
to state $|\Psi^-\rangle$. 
To verify the state  $|\Psi^-\rangle_{1'3'}$, we have to make a collective measurement
which can be carried out through using a beam splitter\cite{bou,sl}.
As it was shown in Ref\cite{bou}, if beam 1' and 3' each contains one photon and 1'' and 3'' are proven to
contain one photon in each beam, then 1' and 3' must be in the singlet state. 
However this is only true in the case that beam 1' and 3' each contains
one photon. Before using this conclusion, we have to carefully study all possible states in beam 1' and 3' 
and the consequences of each of them.
The  state $|\chi\rangle_{1'2'}\otimes |\chi\rangle_{3'4'}$ contains a number of components (i.e., $X_0,X_1,A,B,C$ and higher order terms) with different 
probability amplitude, i.e., the total state is the linear superposed state of those components.
 Let's first study  what happens to each of those different components in the state 
$|\chi\rangle_{1'2'}\otimes |\chi\rangle_{3'4'}$.

 The component with largest probability is
$$|V\rangle_{2'}|H\rangle_{2'}|V\rangle_{4'}|H\rangle_{4'}.$$  The prior
probability( the probability before we make an observation on the photon detectors)
for this component is $P_1=\frac{1}{(1+|\epsilon|^2)^4}$. 
For this component, there is no photon  in beam 1' or 3', therefore no photon
detector will be clicked. Such a component will be definitely ruled out by the conditions
in our coincidence. 

The component with the probability  amplitude order of $\epsilon$ is
$$
(|V\rangle_{1'}|V\rangle_{2'}-|H\rangle_{1'}|H\rangle_{2'})
|V\rangle_{4'}|H\rangle_{4'}+(|V\rangle_{3'}|V\rangle_{4'}-|H\rangle_{3'}|H\rangle_{4'})|V\rangle_{2'}|H\rangle_{2'}.
$$ The prior probability for this component is 
$P_2=\frac{4|\epsilon|^2}{(1+|\epsilon|^2)^4}$.
This means that there is only
one photon altogether in both of beam 1' and beam 3'. One can easily find this fact by checking each term 
of the above formula. Each term only 
allows one photon in beams 1' and 3'.  This component will cause only one detector to be clicked and 
all the other three are silent. Such an event is obviously
different from our required coincidence therefore the above component can
be safely excluded once a coincident event is observed. 

Now we consider the components with the  probability amplitude order of $\epsilon^2$. These components are $A=|H\rangle_{1'}|V\rangle_{1'}|V\rangle_{4'}|H\rangle_{4'}$, $B=|V\rangle_{2'}|H\rangle_{2'}|H\rangle_{3'}|V\rangle_{3'}$ and
\begin{eqnarray}
C=(|H\rangle_{1'}|H\rangle_{2'}-|V\rangle_{1'}|V\rangle_{2'})
(|H\rangle_{3'}|H\rangle_{4'}-|V\rangle_{3'}|V\rangle_{4'})
\end{eqnarray}
The total prior probability for these three components is 
$P_3=\frac{6|\epsilon|^4}{(1+|\epsilon|^2)^4}$.
We now show that 
component $A,B$ will never cause the defined coincident event.
Before we go into that, we first take a look at the properties of the beam splitter  used in our scheme. 
The property of a balanced beam splitter is sketched in figure 3. A detailed study of the properties of
a beam splitter can be seen e.g., in ref.\cite{campos,wang0}.
For clarity,  we use the {\bf Schrodinger} picture here. The different modes are
simply distinguished by the propagation directions. In our case, states of beam 1' and 1''
are of the same mode (we denote it as mode $a$ ) at different times, and beam 3' and 3'' are in 
another mode, mode $b$. Suppose the input beams (1' and 3') are in the state
$|input\rangle$, then the output state (in beam 1'' and 3'') is 
$|output\rangle=U_B|input\rangle$, where $U_B$ is the time evolution operator for the
beam splitter. The unitary operator $U_B$ satisfies
\begin{eqnarray}\label{b1}
U_B (a_H^\dagger,b_H^\dagger,a_V^\dagger,b_V^\dagger) U_B^{-1}
= (a_H^\dagger,b_H^\dagger,a_V^\dagger,b_V^\dagger)\left(
\begin{array}{cc}{\bf H} & {\bf O}\\ {\bf O} & {\bf H}\end{array}\right)
\end{eqnarray}    
where $a^\dagger,b^\dagger$ are creation operators for mode $a$ and mode $b$ respectively, the subscripts $H,V$ indicate the horizontal and vertical
polarizations respectively,  
\begin{eqnarray}\label{hada}
{\bf H}=\frac{1}{\sqrt 2}\left(\begin{array}{cc}1&1\\1&-1\end{array}\right)
,\end{eqnarray}
and ${\bf O}$ is a $2\times 2$ matrix with all elements being 0.
Note that the evolution operator $U_B$ also satisfies
\begin{eqnarray}
U_B|00\rangle_{in}=|00\rangle_{out}
\end{eqnarray}
due to the fact of no input no output. Here the subscripts $in$ and $out$
indicate the input beams and output beams respectively. In our case, the input beams are 1' and 3' the output beams are 1'' and 3'', therefore $|00\rangle_{in}=|00\rangle_{1'3'}$ and  $|00\rangle_{out}=|00\rangle_{1''3''}$. 

With the above two equations, in general the state in the output beams and the state in the input beams are
simply related by:
\begin{eqnarray}\label{b7}
|output\rangle_{out} = 
U_B f(a_H^\dagger,a_V^\dagger,b_H^\dagger,b_V^\dagger)
U_B^{-1}\cdot U_B|00\rangle_{in}
=\left(U_B f(a_H^\dagger,a_V^\dagger,b_H^\dagger,b_V^\dagger)U_B^{-1}\right)|00\rangle_{out}
\end{eqnarray}
provided that 
$|input\rangle_{in}=f(a_H^\dagger,a_V^\dagger,b_H^\dagger,b_V^\dagger)
|00\rangle_{in}$. Note that $U_B f(a_H^\dagger,a_V^\dagger,b_H^\dagger,b_V^\dagger)U_B^{-1}$
can be easily calculated by using eq.(\ref{b1}). In our treatment, all the creation operators are time independent since we 
are using the Schrodinger picture.
Now we  consider the component $A$. Since $A$ is a product state of different modes,
we only consider evolution to the part in beam 1' and beam 3'. 
There is no nontrivial change in beam 2' and 4'.
For the input of component $A$, the  output state
of the BS here is
\begin{eqnarray}
|output\rangle_{1''3''}=B a_H^\dagger a_V^\dagger B^{-1}|00\rangle_{1''3''}.
\end{eqnarray}
This is a direct consequence of eq.(\ref{b7}). Moreover,
using eq.(\ref{b1}) one can easily obtain
\begin{eqnarray}
|output\rangle_{1''3''}=
\frac{1}{2}[(|HV\rangle_{3''}+|HV\rangle_{1''})
+(|H\rangle_{1''}|V\rangle_{3''}+|V\rangle_{1''}|H\rangle_{3''})].
\end{eqnarray}
The exact form of the term $|HV\rangle_{3''}$( or $|HV\rangle_{1''}$) is
 $|0\rangle_{1''}|HV\rangle_{3''}$( or $|HV\rangle_{1''}|0\rangle_{3''}$), it
means beam 3''(or 1'') contains one horizontally polarized photon
 and one vertically polarized photon while beam 1''(or 3'')
contains nothing. The term  $|HV\rangle_{3''}$  causes neither D$_1$ nor D$_2$ being clicked 
therefore   it never causes our defined coincident event is. Similarly,  the consequence
of $|HV\rangle_{1''}$ is that neither D$_3$ nor D$_4$ will be clicked therefore
this term is also ruled out. The term 
$|H\rangle_{1''}|V\rangle_{3''}+|V\rangle_{1''}|H\rangle_{3''}$ means 
that beam 1'' and 3'' each contain one photon. However, after the two half wave plates(HWP) the term is changed to
\begin{eqnarray}\label{half}
|H\rangle_{\alpha}|H\rangle_{\beta}-|V\rangle_{\alpha}|V\rangle_{\beta}.
\end{eqnarray}
One may easily check this result by using the time evolution operator of
the HWP defined as
\begin{eqnarray}
U_H\left(\begin{array}{c}|H\rangle\\|V\rangle\end{array}\right)={\bf H}
\left(\begin{array}{c}|H\rangle\\|V\rangle\end{array}\right).
\end{eqnarray}
Note that {\bf$H$} is defined by eq.(\ref{hada}). 
 Obviously, the two detectors clicked by the state in equation(\ref{half}) 
will be either (D$_1$, D$_3$) or (D$_2$, D$_4$), neither of them is
 the our defined {\bf coincidence }. The coincident event will never happen with the state of eq.(\ref{half}).
Therefore component $A$ is now totally ruled out. 

Similarly, for the component $B$, the output state of the BS is
\begin{eqnarray}
|output\rangle_{1''3''}=B b_H^\dagger b_V^\dagger B^{-1}|00\rangle_{1''3''}
=\frac{1}{2}[(|HV\rangle_{3''}+|HV\rangle_{1''})
-(|H\rangle_{1''}|V\rangle_{3''}+|V\rangle_{1''}|H\rangle_{3''})].
\end{eqnarray}
It's easy to see that component $B$
should be  also ruled out due to the same arguments used in the case of component $A$. 

Now the only component with the same
magnitude order of probability amplitude $\epsilon^2$ is the component $C$:
\begin{eqnarray}\label{cc}
C= |H\rangle_{1'}|H\rangle_{3'}|H\rangle_{2'}|H\rangle_{4'}+
|V\rangle_{1'}|V\rangle_{3'}|V\rangle_{2'}|V\rangle_{4'}-
|\Psi^+\rangle_{1'3'}|\Psi^+\rangle_{2'4'}-
|\Psi^-\rangle_{1'3'}|\Psi^-\rangle_{2'4'})
\end{eqnarray} 
As it is well known, to a beam splitter, if each of the input beam contains one photon
and the total input polarization state is symmetric, one output beam must be vacant.
For the component $C$, each of beam 1' and 3' always contains one photon. The first three terms are all
symmetric states. Given these three terms as the input, one output beam 
of the BS must be vacant,  i.e., either beam 1'' or beam 3'' must be empty. .  
Consequently, given any of the first three terms in component $C$ as the input, one will observe that
 either both (D$_1$, D$_2$) or both  (D$_3$, D$_4$) are silent.  
This definitely violates 
the conditions of our required coincidence therefore 
 the first three terms of the right hand side in eq.(\ref{cc}) 
are excluded for a coincident
event. However, the last term in eq.(\ref{cc}) exactly satisfies the conditions
of our required coincidence. For the input state $|\Psi^-\rangle_{1'3'}$, 
the output state of BS is still a singlet state, 
i.e. $|\Psi^-\rangle_{1''3''}$\cite{bou,sl}. This state is invariant under the transformation of two
separate HWPs. Therefore finally, after the beams pass through the two separate HWPs, the two photons are still in the state $|\Psi^-\rangle$ consequently the required coincidence is observed because the polarizing beam splitters PBS3 and PBS4 evolve the  state $|\Psi^-\rangle$ 
into \begin{eqnarray}
\frac{1}{\sqrt 2}(|x\rangle|w\rangle-|y\rangle|z\rangle).
\end{eqnarray} Here 
$|x\rangle$,$|w\rangle$, $|y\rangle$ and $|z\rangle$  represent 
for the state of one photon in beam $x$, $w$, $y$ and $z$, respectively.
  As we have shown, in all terms with the same probability amplitude order, state $|\Psi^-\rangle_{1'3'}|\Psi^-\rangle_{2'4'}$ is the only term that causes our defined coincident event. Therefore once a coincident event is observed, beam 2' and 4' must be in the state
$|\Psi^-\rangle_{2'4'}$ with a probability close to 1\cite{noteclose}.

In our scheme,
 the total probability that a coincident event takes place is around 
$|\epsilon|^4$.  In the above study, we have ignored the effects of those components
with a probability amplitude order higher than  $|\epsilon|^2$.  Whenever a 
coincident event is observed, although 
the state $\rho_{2'4'}$ for the outcome beams is very close to the singlet state $|\Psi^+\rangle_{2'4'}$, it is not the 
perfectly pure singlet state because the outcome
beams could be a single photon state or a vacuum state with a very small probability (the magnitude order of
probability amplitude is $\epsilon^3$). To calculate  
the fidelity between the produced state $\rho_{2'4'}$  and the perfect singlet state $|\Psi^-\rangle$,we need calculate the post probability (the probability after the observation of the coincident events) of $\rho_{2'4'}$ being the singlet state. In the case that a coincident event takes place, the component
$A$,$B$, the first 3 terms in component $C$ and all of the 
terms with the probability
amplitude order lower than $|\epsilon|^2$ are excluded. The 
prior probability of all those excluded states is
\begin{eqnarray}
P_{im}=\frac{1}{(1+|\epsilon|^2)^4}(1+4|\epsilon|^2+5|\epsilon|^4). 
\end{eqnarray}
To calculate the lower bound of the fidelity, we assume the worst situation 
that all omitted higher order terms will cause the coincident events. 
In such a situation, when a coincident event is observed, the fidelity between $\rho_{2'4'}$ and the singlet 
state is 
\begin{eqnarray}
\langle \Psi^-|\rho_{2'4'}|\Psi^-\rangle 
\ge\frac{|\epsilon|^4/(1+|\epsilon|^2)^4}{1-P_{im}}
= \frac{|\epsilon|^4}{(1+|\epsilon|^2)^4-(1
+4|\epsilon|^2+5|\epsilon|^4)}=1-4|\epsilon|^2.
\end{eqnarray}
This is to say, if we set $\epsilon=\frac{1}{20}$, we can make a singlet state
 in beam 2' and 4' with a purity larger than $99\%$, once the coincident event is observed.
 Note that this is the lower bound of the fidelity, since
we have assumed all the terms with probability amplitude order higher than
$|\epsilon|^2$ will cause a coincident event wrongly. Actually, some of the
higher order terms will not cause the required coincidence by our scheme and this will increase the actual fidelity. A detailed calculation shows that the actual fidelity is larger than $99.7\%$.

In general, the efficiency of a photon detector is far from perfect. In our scheme, if the efficiency is $\eta$, 
the total probability that a coincident event takes place is changed to $\eta^2|\epsilon|^4$. When a coincident
event happens, the lower bound
of the purity of the outcome beams is $1-4|\epsilon|^2\eta^{-2}$. This is to say, e.g. given the efficiency
$\eta=0.5$ and $\epsilon=\frac{1}{20}$, the fidelity between $\rho_{2'4'}$ and the singlet state is larger
than $94\%$. Again, this value is only the lower bound of the fidelity.  A detailed calculation 
shows that the actual fidelity will be larger than $99\%$. 
 
In conclusion, by using the scheme as shown in figure 1, we can prepare 
 a good EPR state in beam 2' and 4' conditionally on the  observation of  the coincident
event that both D$_1$ and D$_4$ are clicked or both D$_2$ and D$_3$ are clicked. As far as we 
have known, so far this is the $only$ passive
linear optics scheme to produce
EPR pairs with the normal photon detectors which only distinguish the vacuum
and non-vacuum Fock state. It has already been shown\cite{single2} that the single photon 
state can be produced successfully with nearly $100\%$ probability with the pump light
repetition rate of $10^8$ per second, by using the quantum dot technique. Moreover, the single photon state
can be mass produced by the robust electrically driven source\cite{single4}. The synchronization
of two beams of single photon is a challenging task in practice. But it is not an unsolvable 
problem in principle. Actually, the indistinguishability and the interference of two single photon
beams have been indeed experimentally observed recently\cite{single5} though the experimental
efficiency there is low. 
As it has been calculated earlier, even imperfect normal photon detectors
 with an  efficiency 
$50\%$ are used, the fidelity between our outcome state and the perfect EPR state is still quite high. This
is a bit different from the third order SPDC scheme given by Sliwa and Banaszek\cite{bana} where the
result is seriously distorted by the low detection efficiency.

{\bf Acknowledgement:} I thank Prof Imai H for support. I thank Dr Matsumoto K,  Tomita A and  Pan JW( U. Viena) for stimulating discussions. 

\newpage
\begin{figure}
\begin{center}
\epsffile{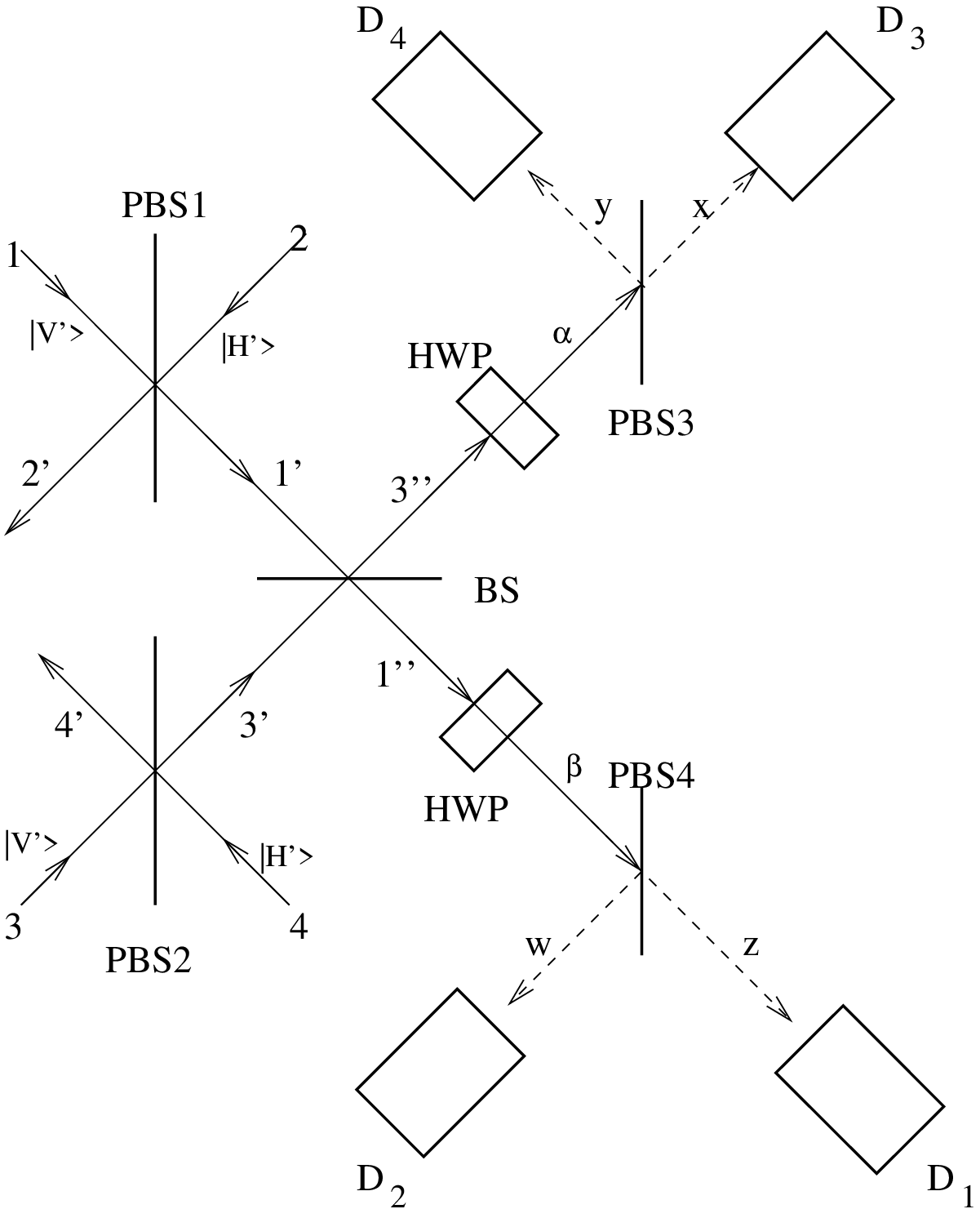}
\end{center}
\caption{A schematic diagram for our scheme to produce the polarization
entangled state with linear optics devices.
The input  beams 1,2,3,4 are  from independent single photon sources.
Initially, the polarization of beam 1 and the polarization of beam
 3  deviate a little bit from
the vertical one; the polarization of beam 2 and the polarization of beam
 4  deviate a little bit from the horizontal
one.
If the detector D$_1$ and D$_4$ are both clicked, or if the detector
D$_2$ and D$_3$ are both clicked, a state $\rho$ which is very close to the singlet state $|\Psi^-\rangle_{2',4'}$ has been prepared in beam 2' and 4'. For all the beams as the input of a polarizing beam splitter(PBS) or a beam splitter(BS), the optical paths should be arranged carefully to make sure the 
two input beams reach a PBS or a BS in the $same$ time.}
\end{figure}
\begin{figure}
\begin{center}
\epsffile{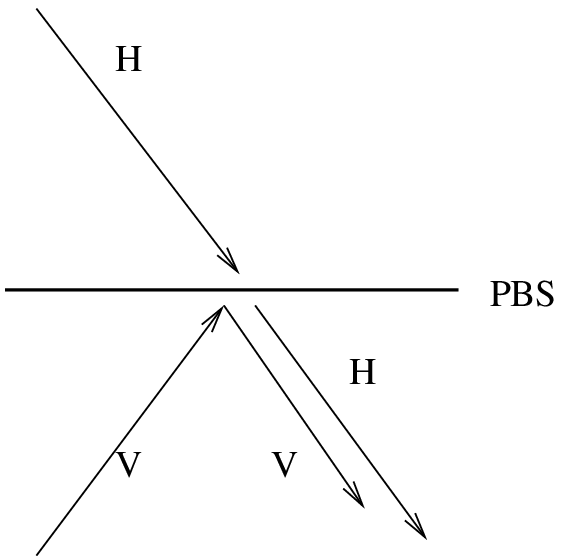}
\end{center}
\caption{ A schematic diagram for the property of a polarizing beam spliter(PBS). 
It transmits a horizontally polarized
photon $H$ and reflects a vertically polarized photon $V$.} 
\end{figure}
\begin{figure}
\begin{center}
\epsffile{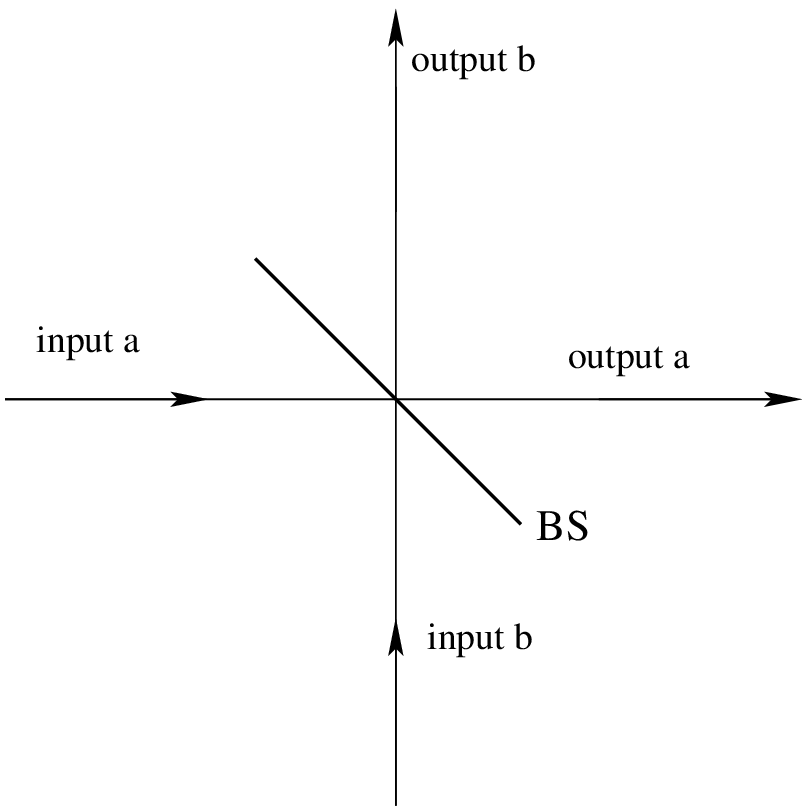}
\end{center}
\caption{ A schematic diagram for the beamsplitter operation. Both the input and the output
are two mode states. The different mode is distinguished by the propagating direction of the 
field. If the input state is $f(a^\dagger_H,a^\dagger_V,b^\dagger_H,b^\dagger_V)|00\rangle$, the output state is 
$U_Bf(a^\dagger_H,a^\dagger_V,b^\dagger_H,b^\dagger_V)|00\rangle=U_Bf(a^\dagger_H,a^\dagger_V,b^\dagger_H,b^\dagger_V)
U_B^\dagger|00\rangle$}, where $U_B$ is the time evolution operator of the beam splitter. 
\end{figure}

\begin{thebibliography}{99}
\bibitem{remark1} In the Innsbruck experiment\cite{bou}, the polarized weak Gaussian beam created by the
spontaneous parametric down conversion process is used to replace the maximum
polarization entangled state. This polarized weak Gaussian beam can be  regarded as the probabilistic 
two photon EPR state. Consequently, the teleportation task there can only be done sucessfully
with a small probability.  Given the event-ready entanglement, the goal of deterministic quantum teleportation can be done.
\bibitem{noteclose} The terms with probability amplitude order of $\epsilon^3$ and $\epsilon^4$ will
distort the outcome a little bit.
\end{thebibliography}
\end{document}